# Conceptual Modeling with Constraints

Sabah Al-Fedaghi

*salfedaghi@yahoo.com, sabah.alfedaghi@ku.edu.kw*
Computer Engineering Department, Kuwait University, Kuwait

## Abstract

An important factor in guaranteeing the quality of a system is developing a conceptual model that reflects the knowledge about its domain as well as knowledge about the functions it has to perform. In software engineering, conceptual modeling has gained importance as a discipline that offers languages, methods, and methodologies to address the complexity of software development. The key to understanding such complexity is using tools such as diagrams at various levels of representation. A conceptual model must include all relevant static and behavioral aspects of its domain. In UML, the static aspects include structural diagrams that represent the internal architecture of a system with a special focus on the classes, the connections and interactions that they have, and integrity constraints over the state of the domain. UML does not have sufficient expressiveness for complete specifications of certain constraints. Constraints assist in analyzing permissible design requirements and the limitations of the intended functions. To overcome the limitations of the graphical notation, other types of languages are used to complement the diagrammatic language (e.g., the textual Object Constraint Language [OCL]). In this paper, we study how to express constraints diagrammatically using the thinging machine (TM) through examples taken from the UML/OCL literature. This would contribute to further understanding the notion of constraint in conceptual modeling. It also demonstrates the expressiveness and limitation of the TM. The paper suggests that the TM can provide a diagrammatic constraints language in conceptual models.

## 1. Introduction

Software development is said to be one of the most challenging engineering activities [1][2]. According to Sommerville [3], this is why software modeling has gained importance as a discipline that offers languages, methods, and methodologies to address the complexity of software development. The key to understanding of such complexity is the use of tools such as diagrams at various levels of representation. In this context, modeling addresses, among other things, the quality of a system by developing a conceptual model that reflects the knowledge about its domain as well as knowledge about the functions it has to perform [4].

Conceptual modeling must include all relevant static and behavioral aspects of its domain [5]. The static aspects include structural diagrams that represent the internal architecture of a system with a special focus on the classes, the connections and interactions that they have, and integrity constraints over the state of the domain, which define conditions that each state of the modeled system must satisfy [6]. According to Jairo [7], the structural description does not detail the internal logic, only the inputs and outputs.

Behavioral aspects of a system refer to operations and the definition of their effect, including the changes they make when they are executed. These dynamics are usually specified by means of a behavioral model consisting of a set of system operations [8]. A representative diagram of the behavior is defined as a diagram that represents the different states of the process. Behavioral description must be consistent with regard to structural specification. Consistency refers to not having a contradiction or unsatisfiable structural entities, e.g., classes.

### 1.1 UML/OCL

The Unified Modeling Language (UML) provides structural specifications of several diagrams, including the class diagram, the backbone of UML, which is used to define the entity types and relationship types together with some constraints that can be expressed graphically [6]. The constraints include constrained elements—association classes and aggregation/composition properties—together with other constraints on these elements, including cardinality constraints on properties and attributes, class hierarchy constraints, generalization set constraints, and inter-association constraints [9].

Nevertheless, the class diagram is not sufficient for a precise and unambiguous specification about the objects in the model; hence, there is a need to describe additional constraints. Some constraints that cannot be expressed graphically can be expressed by means of the Object Constraint Language (OCL) [10]. The OCL is a formal high-level language used to write expressions on UML models. It is a textual language that provides constraint that cannot otherwise be expressed by diagrammatic notation [11]. The OCL has been extended to include general object query language definitions.

## 1.2 Constraints

According to Kamarudin et al. [12], constraint is the key to understanding complexity. A constraint-based problem can spark ideas for new knowledge, new possibilities, and new opportunities. In every design, boundaries, controls and restraints exist. A rule is a law or regulating principle for producing a certain result or solution. A constraint is a restriction or a condition that a lawful solution to a problem must satisfy. It is a limitation under which a system must operate, e.g., cost, time etc. In conceptual modeling, constraints should be the first to be studied since they assist in analyzing permissible design requirements and the limitations of the functions' work together. Inappropriate constraint management in conceptual design can cause catastrophic failure, but removing constraints will result in a chaotic system [13].

## 1.3 Problem: Validation of structural and behavioral diagrams

A UML/OCL model and its constraints should be validated and verified before the start of its implementation because many design mistakes and implementation faults can thus be avoided [14]. Validation (i.e., finding out inconsistencies) and verification of UML class diagrams constrained by OCL invariants is an open question of research and a topic of great interest [14]. According to Mokhtari [15], (2020), there is no way to verify the satisfaction of the OCL constraint properties by a modeled system. The UML class diagram analysis is a complex problem. The addition of OCL constraints makes the problem unsolvable in general [16].

According to Khan and Porres [17], although a lot of research work has already been done in the area of the validation of structural and behavioral diagrams, there is room for new approaches in this area. OCL is just a special case of a general pattern where diagrammatic modeling languages use textual languages to define constraints that are difficult to express with their own syntax and semantics. The identification of classes of modifications for which an automatic synchronization of OCL constraints is possible requires tool vendors to implement a complex machinery [18]. Any modification in a model structure (e.g., UML) must be reflected in the OCL constraints which are related to the modified structure, but defining automatic synchronization of OCL constraints for arbitrary model modifications is not possible [19].

In general, the OCL is a language whose spread has not met the optimistic expectations expressed since its inclusion as part of UML 1.1 due to the ambiguities and gaps in the language specification. According to Chiorean et al. [20], "Although there has been some progress in the above mentioned fields, the developers' feedback is far from satisfactory." According to Queralt Calafat [6], due to the high expressiveness of the combination of the UML and OCL languages, checking the correctness of a UML conceptual model manually becomes a very difficult task, especially when the set of textual constraints is large. There are a few proposals that take the behavioral model into account in the validation process, none of them dealing with UML schemas with general OCL constraints and operations.

## 1.4 About this paper

In this paper, we study how to express constraints diagrammatically using the thinging machine (TM) through examples taken from the UML/OCL literature. This will contribute to greater understanding of the notion of constraint in conceptual modeling. The paper also demonstrates the expressiveness and limitation of the TM. We propose that the TM can provide a diagrammatic constraints language in conceptual models.

The next section presents a brief description of the TM with an example. The remaining part of the paper presents re-modeling of examples from the literature that involve constraints.

## 2. Thinging Machine (TM)

Models can be viewed as frameworks for organizing knowledge without presupposition that models must resemble the real world in any form or fashion [21]. The TM (see [22][23]) views the world as a thimac (things/machines) constructed from thimacs. The thimac is an encapsulation of a thing that reflects the unity and hides the internal structure of the thimac, and a machine (see Fig. 1) shows the structural components (called region), including *potential* actions of behavior. The static "thing" does not actually exist, change or move, but it has potentialities for these actions when combined with time. A TM event is an encapsulation of a region and time.

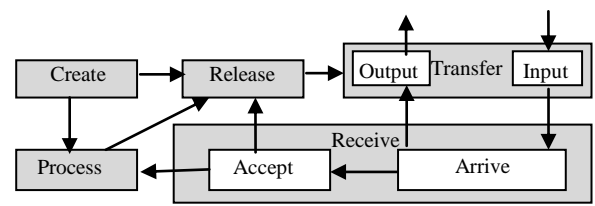

Fig. 1 Thinging machine.

## 2.1 Things that are machines

A thimac is a thing. A thing is what can be created (seen, observed), processed (changed), released, transferred, and/or received. A thing is manifested (can be recognized as a unity) and related to the "sum total" of a thimac. The whole TM occupies a conceptual "space" that forms a compositional structure of thimacs that link together like the links of a whole network. The whole is a grand thing/machine. Thimacs can be "located" only via flow connections among thimacs. The thimac is also a machine that creates, processes, releases, transfers, and/or receives. Fig. 1 shows a general picture of a machine. The figure indicates five "seeds" of potentialities of dynamism: creation, processing, releasing, transferring, and receiving.

All things are created, processed, released, transferred, and received, and all machines (thimacs) create, process, release, transfer, and receive other things. Things "flow through" (denoted by a solid arrow in Fig. 1) other machines. Thus, things flow within other things. Thimacs flow through other thimacs. A TM event is a thing comes into being when a region (subdiagram) combines with time. This picture is in line with the Heraclitean idea that to be alive is to inhale something new and mysterious [24]. The thing in a TM diagram is a presentation of any "existing" (appearing) entity that can be "counted as one" and is coherent as a unity.

Fig. 1 can be described in terms of the following generic actions (those having no more primitive action):
**Arrive**: A thing moves to a machine.
**Accept**: A thing enters the machine. For simplification, we assume that all arriving things are accepted; thus, we can combine the *arrive* and *accept* stages into one stage, the **receive** stage.
**Release**: A thing is ready for transfer outside the machine.
**Process**: A thing is changed, handled, and examined, but no new thing results.
**Create**: A new thing "comes into being" (is found/manifested) in the machine and is realized from the moment it arises (emergence) in a thimac. Things come into being in the model by "being found."
**Transfer**: A thing is input into or output from a machine.

Additionally, the TM model includes the **triggering** mechanism (denoted by a dashed arrow in this article's figures), which initiates a flow from one machine to another. Multiple machines can interact with each other through the movement of things or through triggering. Triggering is a transformation from one series of flows to another.

A *thimac world* is a very inclusive thing. Anything about that thimac (e.g., a person) and of interest (e.g., in modeling) is to be abstractly included. Likewise the thimac world is inclusive in time (e.g., a living or dead person) and can be part of this same world.

A thimac is connected to its *sub*thimacs (parts) if there is any flow or triggering between them (e.g., the existence (creation) of a person triggers the existence of a body, etc.). Any thimac has storage compartments for its instances. It also interacts with its subthimacs and other outside thimacs through lines of flow knitted by actions.

## 2.2 Example

Before introducing the issue of modeling constraints in the TM, to familiarize readers with TM modeling, we introduce a an example given by the IBM Rational Software Modeler [25] and shown in Fig. 2 that models a class that represents a shopping cart relating to classes that represent customers, purchase orders, and items for sale. To save space, we are going to model only *addItem* and *removeItem* without showing the attributes, as they are understood.

### 2.2.1 TM static model

A class as a thimac is a type of self-contained orderly "world" that includes instances of this world. It contains structure (e.g., subthimacs) and its internal dynamics through five actions that construct (e.g., create) and handle its instances. The totality of thimacs is a thimac. Similarly, properties are small world subthimacs which themselves may be subdivided into even simpler sub-subthimacs.

Fig. 3 shows the corresponding TM static model. There are three machines: Shopping Cart (yellow number 1), Customer (2), and Item (3). The customer gets a shopping cart (pink 4 and 5). The cylinder indicates a collection of carts. Adding Items starts in the Customer (blue 6), where a request for an item is created that moves to *Items* (7) to be processed (8). The processing triggers the retrieval of the data (record) of the ordered item (e.g., it includes price) (9), which flows to the customer (10). The item is directed (11) to a machine (module) called *insert in list* (12) that receives ordered items (13 and 14). The ordered item and the list of already ordered items are processed (15) to trigger (16) the creation of a new list (17) that includes the newly ordered item.

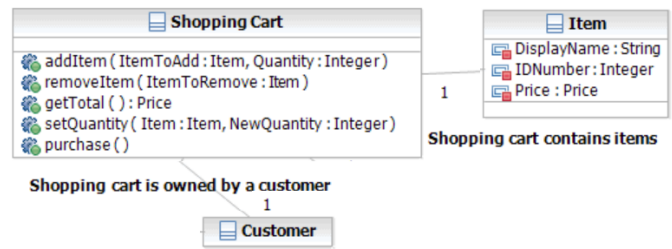

Fig. 2 A class that represents a shopping cart relates to classes that represent customers and items for sale.

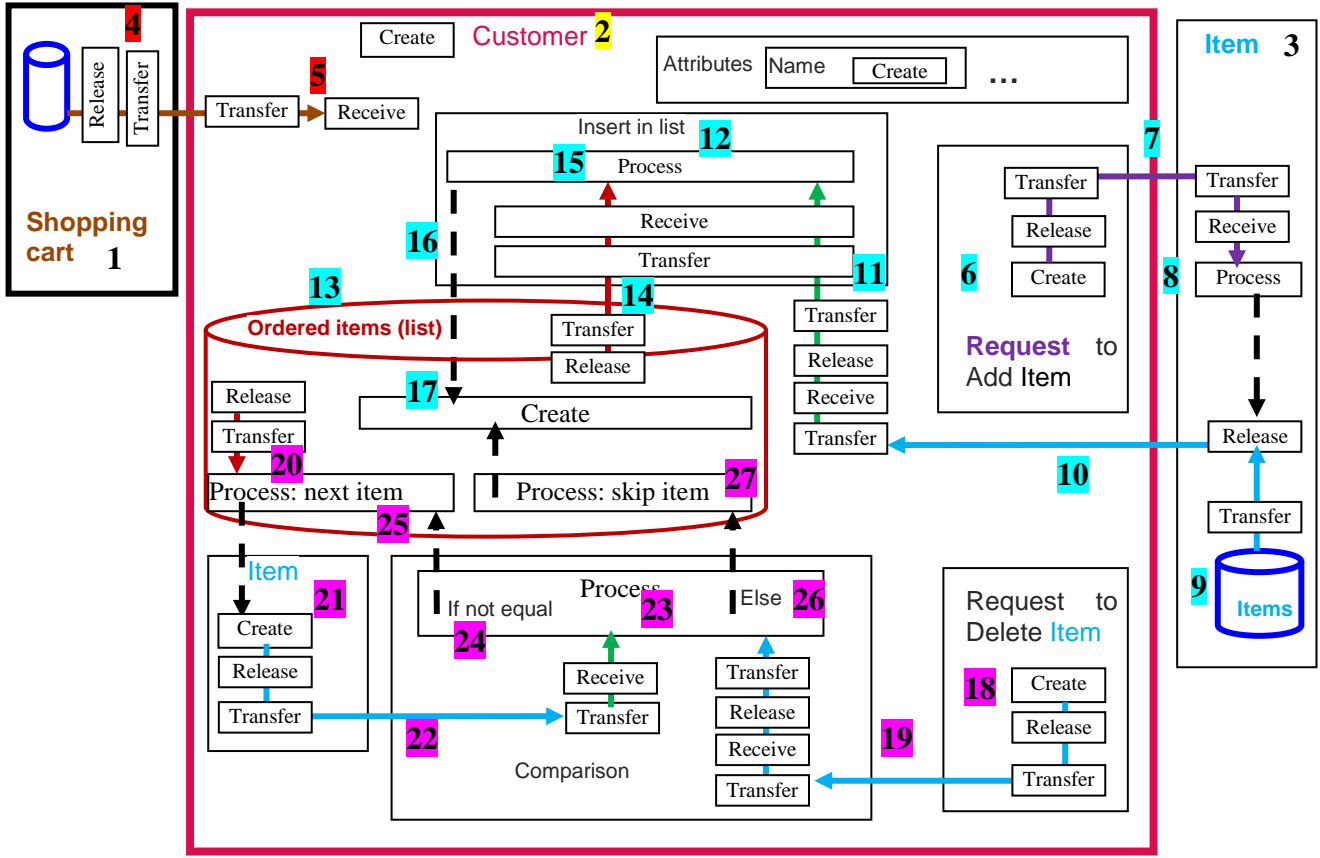

Fig. 3 The TM static model.

To delete an item from the cart involves creating a request to delete the item (pink 18) that flows (19) to a comparison machine (module). For the purpose of this comparison (search for the item in the current list), the ordered list in the cart is processed (20) to extract one item from the list (21). The extracted item moves to the comparison machine (22). The two items are compared (23) and,

- If they are not equal (24), then the next item in the list is extracted (25).
- Else (26), the item in the list is skipped (27) to create (28) a new list without that item.

### 2.2.2 Events and behavior models

An event in TM is a subdiagram (call region) of the static diagram and a time subdiagram. For example, Fig. 4 shows the event *The customer gets a shopping cart*. For simplification's sake, the event may be represented by its region.

Accordingly, Fig. 5 shows four selected events that correspond to the static model of Fig. 3.

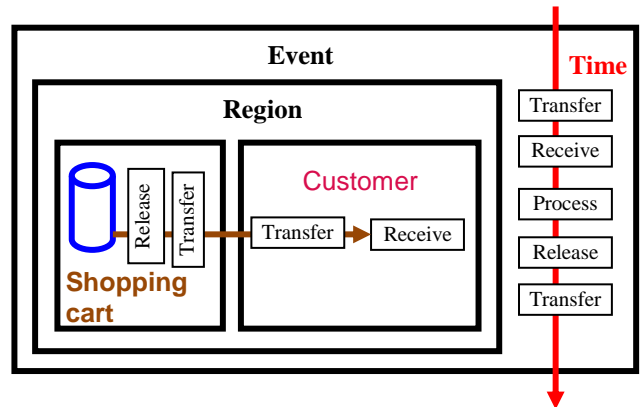

Fig. 4 The event *The customer gets a shopping cart*.

E1: A customer appears in the system, i.e., there exists a customer.
E2: The customer gets a shopping cart.
E3: An item is added to the cart.
E4: An item is removed from the cart.

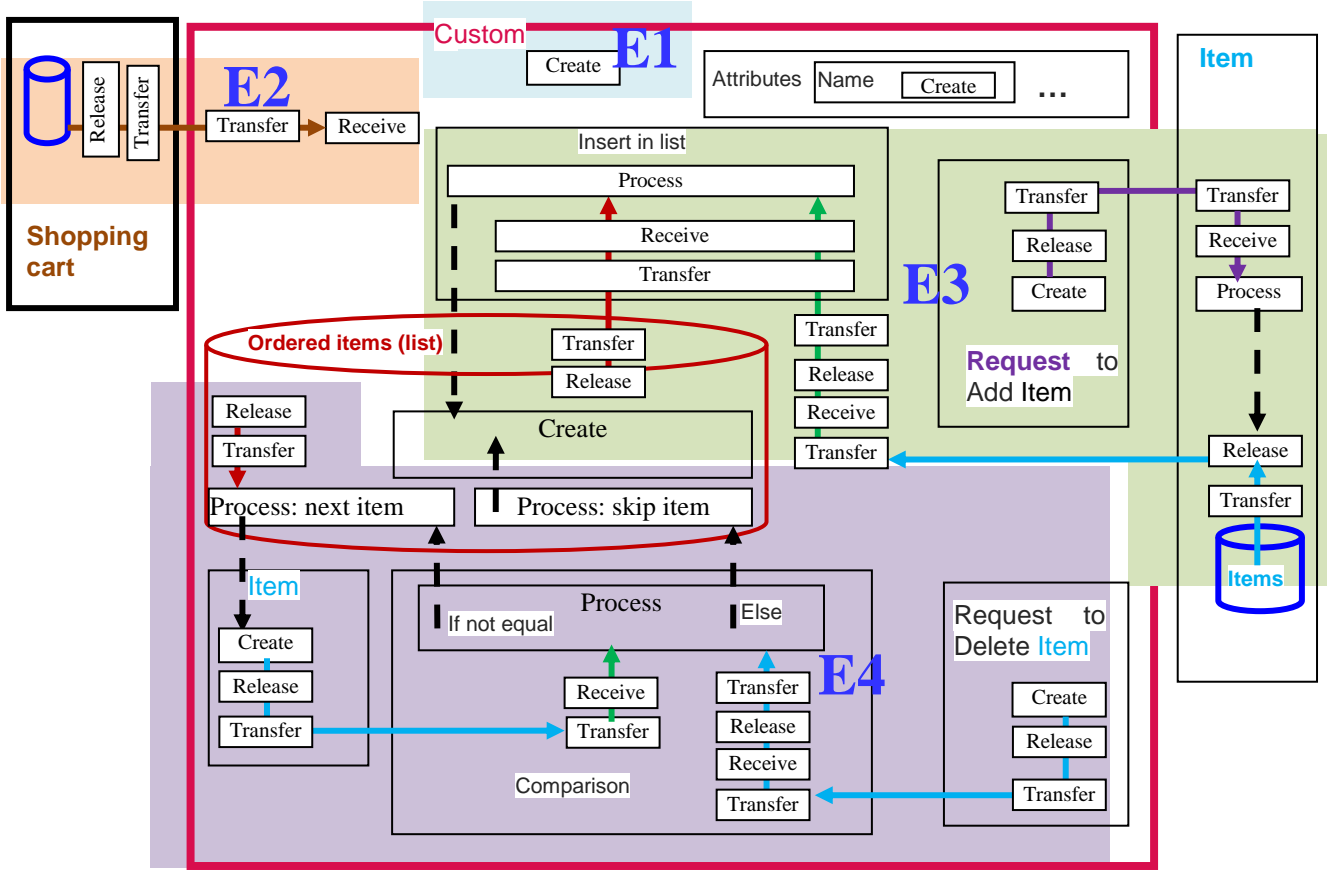

Fig. 5 The events model.

Fig. 6 shows the behavior model. Naturally, this model is a portion of the system. Many functions can be added, such as listing a certain list, checking for deleted item that is not in the ordered list, finishing, etc. Note that the TM diagrams can be simplified. For example, in the static model (Fig. 3), the operations release, transfer, and receive can be eliminated under the assumption that the arrows indicate the direction of flow as shown in Fig. 7.

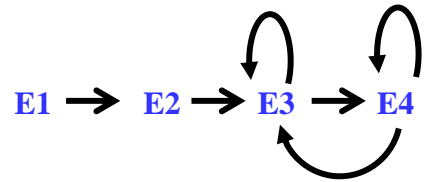

Fig. 6 The behavior model.

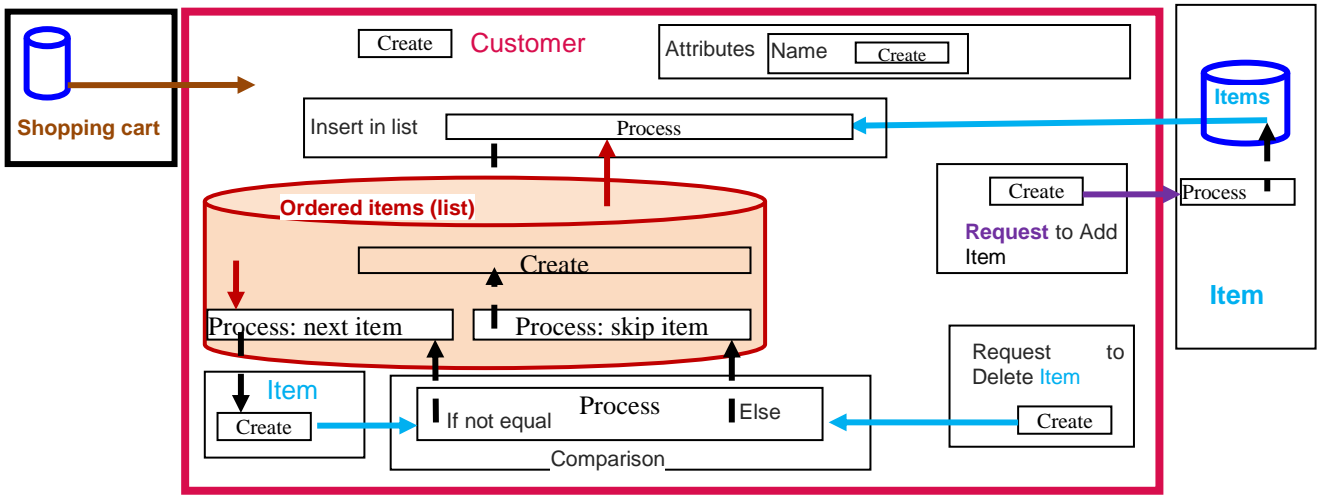

Fig. 7 Simplified TM static model.

## 3. Constraints

As mentioned in the introduction, OCL is a textual language that provides constraints that cannot otherwise be expressed by diagrammatic notation. As an example, according to Warmer and Kleppe [26], the association between the class *Flight* and the class *Person* in Fig. 8, indicating that a certain group of persons are the passengers on a flight, has multiplicity on the side of the *Person* class. In reality, the number of passengers is restricted to the number of seats on the airplane that is associated with the flight. However, it is impossible to express this restriction in the diagram. In this example, the correct way to specify the multiplicity is to add a corresponding OCL constraint to the diagram.

Fig. 9 shows the TM static model that corresponds to this flight class. There are the Person (Pink 1), Flight (2), and Airplane (3). In Person, a name is sent to Flight (Green 4 and 5). Receiving the name (6) triggers processing (7) of the current number of passengers with seats (denoted as x (8)). We assume that x is initially equal to zero. The processing of x increments it, producing y (9). The number of seats of the airplane, n (10 and 11), and y (12) are compared (13).
- If y > n, then a rejection message is produced (14).
- Else, y replaces x as the current number of occupied seats (15) and the name is added to the list of passengers (17 and 18).

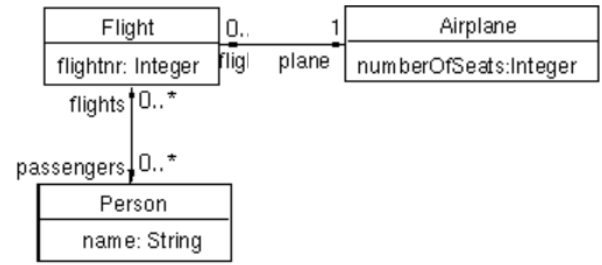

Fig. 8 Flight class model (From [26]).

Fig. 10 shows a simplification of Fig. 9 by eliminating the sequence release, transfer, transfer, and receive, assuming that the direction of the arrow is sufficient to indicate the direction of flow. This simplified model of Fig. 10 can be used to identify selected events as shown in Fig. 11 as follows.

E1: A name is input to be added to the flight.
E2: The number of occupied seats is compared with the total number of available seats.
E3: A seat is available, and thus the passenger name is added to the fight.
E4: No seat is available, hence a rejection message is sent back.

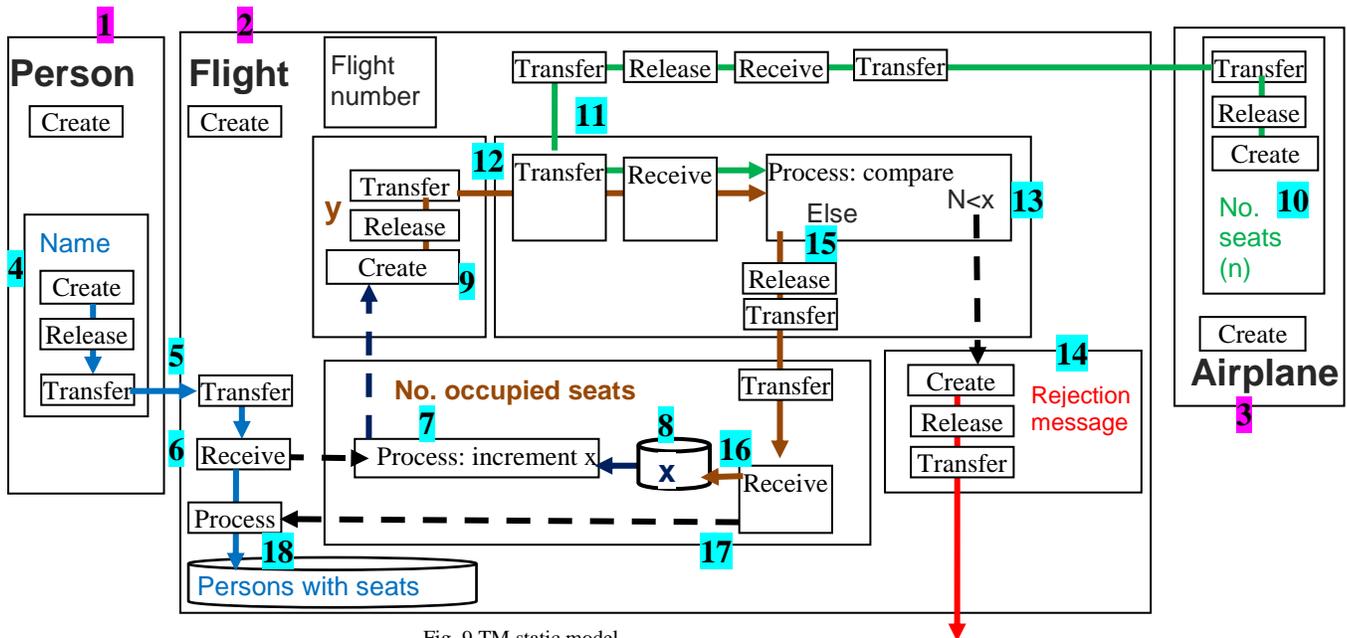

Fig. 9 TM static model.

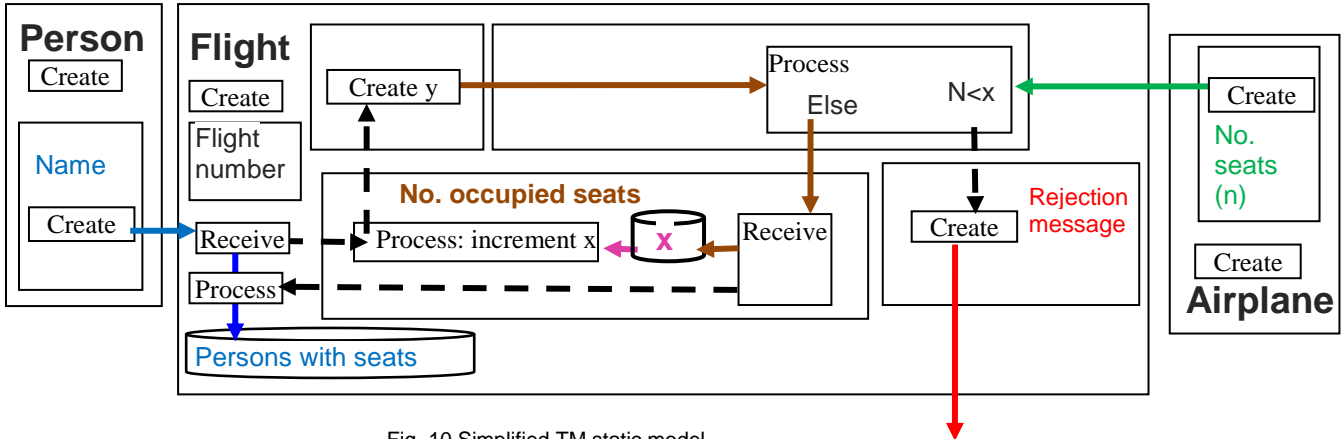

Fig. 10 Simplified TM static model.

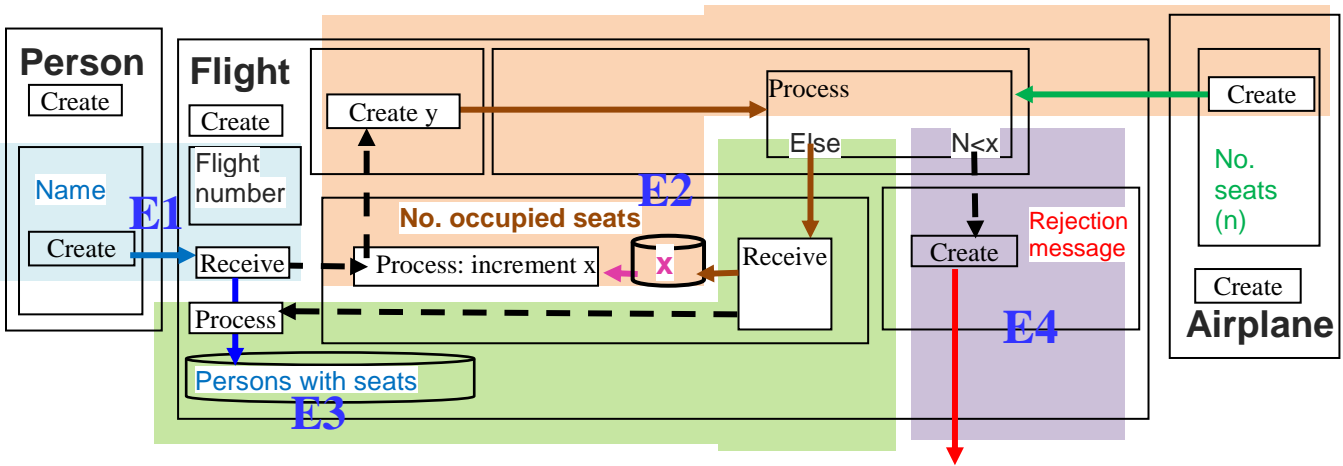

Fig. 11 TM events model.

Fig. 12 shows the TM behavior model. Fig. 13 shows a sample pseudo code script to a sample run.

Accordingly, The TM static model can facilitate the structural description and any declared constraint as demonstrated in this example.

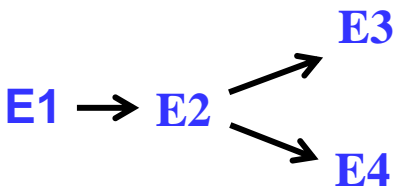

Fig. 12 TM behavior model.

```
Create Airplane=A380   */create an instance of an airplane called A380/*
Create Airplane A380. NoSeats=300        */put the value 300 as the
                                            number of seats of A380/*
Create Flight=Flight1   */ create an instance of Flight called Flight1/*
Create Flight=Flight1.FlightNo=3825        */Put the value 3825 as the
                                            flight number of Flight1/*
Create Person=Person1   */create an instance of Person called Person1/*

E1:
Create.Person=Person1.Name=Michael.release.transfer→Flight=Flight1.
                            FlightNo=3825.Transfer.Receive
Trigger Event E2   */implicitly E2 is applied to "Michael" and "3825"/*
If E3 print "OK"
If E4 print rejection message
```

Fig. 13 Sample script of adding a passenger to a flight.

## 4. Constraints in Logic

In practice, constraints are captured in a natural language such as English and then expressed in OCL. According to Bajwa et al. [27], it is common knowledge that OCL is difficult to write specifically for new users with little or no prior knowledge of OCL. They give the sample constraint; *A customer cannot place more than one order*, modeled in UML and expressed in logic in Fig. 14.

In this section, we construct the corresponding TM model, which is shown in Fig. 15. The TM diagram facilitates expression of the constraint because it is not a pure structural diagram but incorporates potential actions. In Fig. 15, when a customer places an order (number 1 in the figure), e.g., via his/her screen, the order is processed (2). Assuming that the number of orders is initialized to zero, accordingly,

- If the number of orders is > zero, then an error message is issued (3).
- If the number of orders is equal to zero, the order is released (4) and sent to the module order (5).

Naturally, the number of orders is reset to zero when the order is delivered.

Fig. 16 shows the events model, and Fig. 17 shows the behavior model.

## 5. Everything is a model

Rutle et al. [18] introduced a formal diagrammatic approach to modeling based on category theory. According to the authors, an *appropriate* approach to object-oriented modeling is to describe models as graphs. However, the expressive power may not be sufficient to represent certain constraints a system must obey. Accordingly, the authors investigated a *completely* diagrammatic approach for the specification of structural models. This approach obeys the "everything is a model" rule by having both structure and constraints in the same model-centric format. They give an example that illustrates the usage of some constraints which are not expressible by UML itself. These constraints are specified in OCL. In Fig. 18, from Rutle et al. [18], a UML class diagram of an information system for the management of employees and projects is presented. Rutle et al. [18] require that the following set of rules be satisfied at any state of the system:

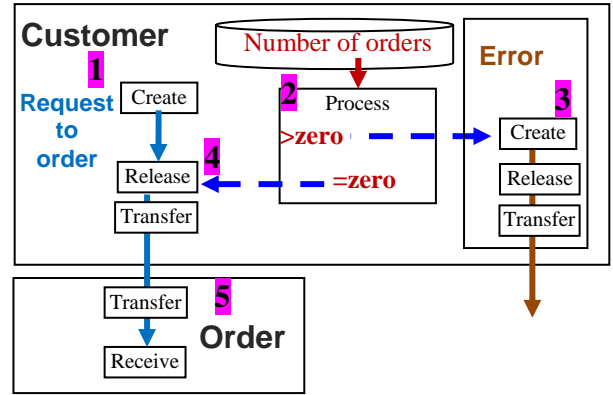

Fig. 15 The TM static model.

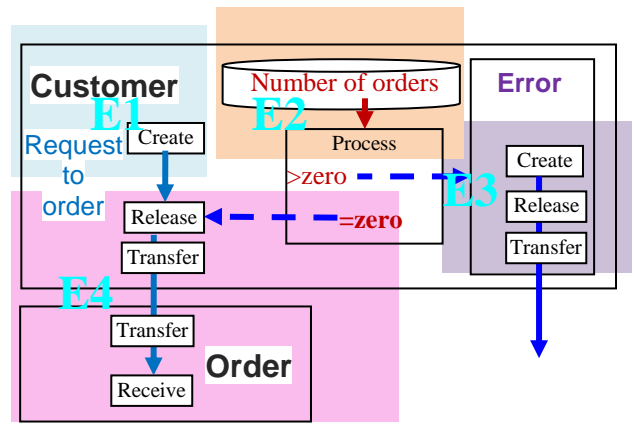

Fig. 16 The TM events model.

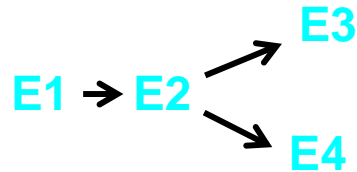

Fig. 17 The TM behavior model.

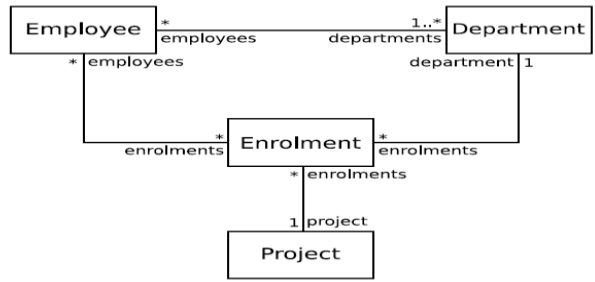

Fig. 18 A UML class diagram for the management of employees and projects [18].

| English: | A customer can place one order. |
|---|---|
| Semantic Interpretation: | |

( place
        (object_type = (∀X ~ (customer ? X)))
        (object_type = ($\exists_{=1}$Y ~ (order ? Y))))

Fig. 14 Sample constraint and its expression in logic (From [27]).

1. An employee must work for at least one department.
2. An employee may be enrolled in no or many projects.
3. A department may have no or many employees.
4. A department may control no or many projects.
5. A project must be controlled by at least one department.
6. An employee enrolled in a project must work in the controlling department.
7. A set of employees working for a controlling department must not be enrolled in the same controlled project more than once.

## 5.1 TM models

Fig. 19 shows the TM static model that corresponds to the given example, according to our understanding. The figure shows only the basic items mentioned by Rutle et al. [18]. There are three thimacs: employee, department, and project.

In the figure, an employee is created (number 1—e.g., a record in the database) and flows to join the appropriate department (2). Similarly, a project (3) is controlled by a department (4). An employee may join a certain project (5). In the department, the number of employees is incremented when an employee joins the department.

Fig. 20 shows selected events in this example. The following events are defined:
E0: A department (i.e., instance) is created.
E1: The number of employees in the department is set to zero
E2: A new employee is created.
E3: An employee joins a department.
E4: The number of employees in the department is incremented.
E5: A project is created.
E6: A project is assigned to a certain department.
E7: An employee is employed in a certain project.

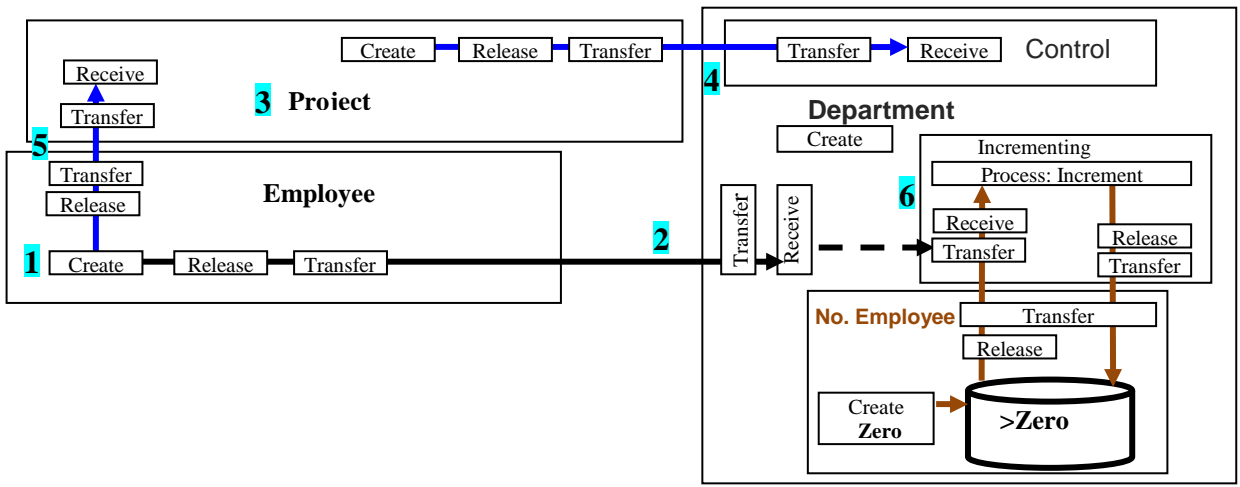

Fig. 19 The static model.

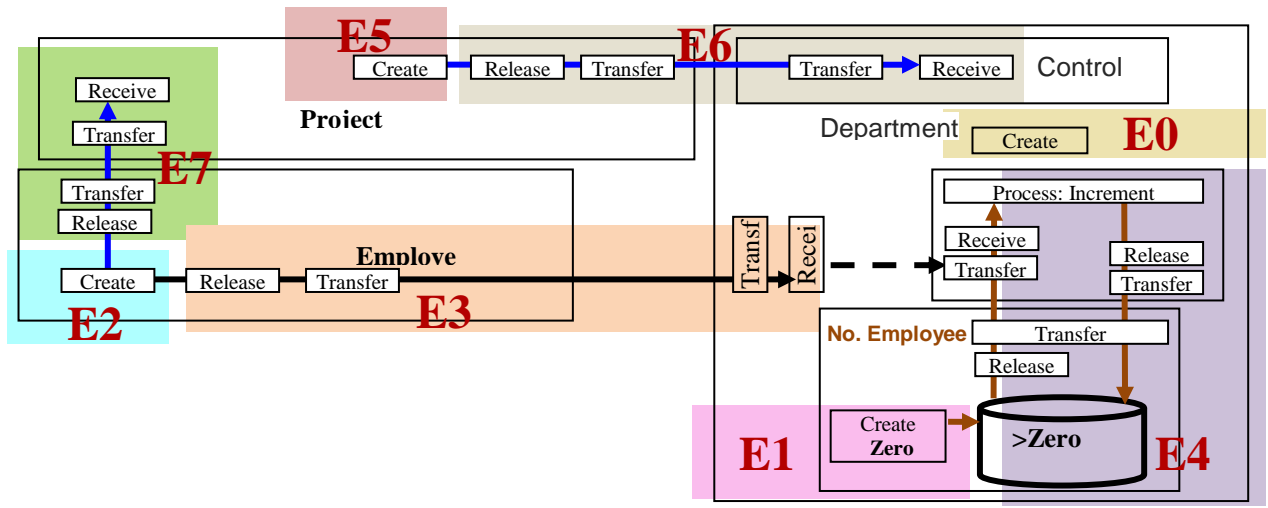

Fig. 20 A behavior model.

Fig. 21 shows the TM behavior model.

## 5.2 Modeling constraints

The basic idea in this section is realizing some constraints by binding events together. Consider the given rules as follows.

**Constraint 1**. An employee must work for at least one department.
Such a rule can be enforced by binding the events E2 and E3 together to form the high-level event **E2-3** as represented by the dotted box in Fig 22. That is,

*When a new employee x is created (E2), then this employee joins department y (E3).*

Thus, no employee exists that is not in a department. The system would require a department name whenever a new employee is created.

Note that E2 may be interpreted as "creating a new x" or "there exists x." Thus, the event E2-3 specifies: if x is a new employee, then create it and then make him/her join department x OR if x already exists (from a previous E2-3), then make him/her join, additionally, department y. Thus, as in the given rule, an employee may join more than one department.

**Constraint 2**. An employee may be enrolled in no or many projects.

This rule is available in the behavioral model since E2 → E7 may be applied several times. In this case the semantics state that an existing employee (E2) *may* join a project (E7).

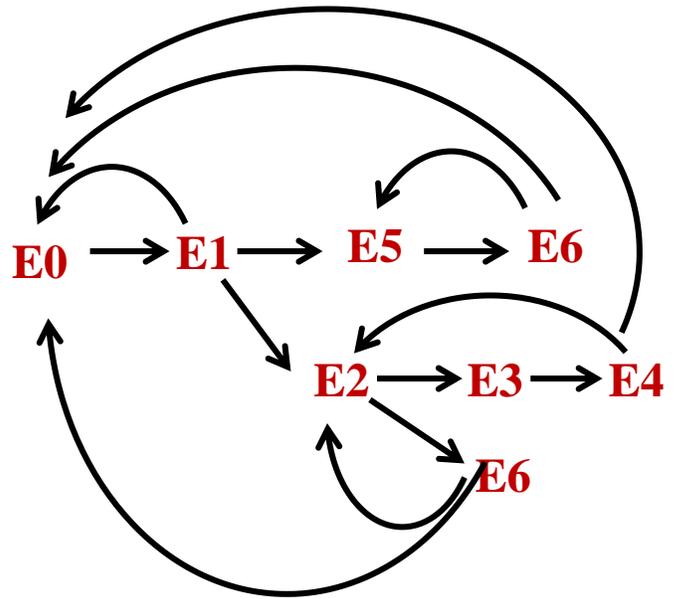

Fig. 21 Eventsmodel.

**Constraint 3**. A department may have no or many employees. This rule is enforced by E4, where a department's number of employees is initialized to zero (E0 is immediately followed by E1 in the behavior model). Additionally, the number of employees is incremented when an employee joins the department (E4).

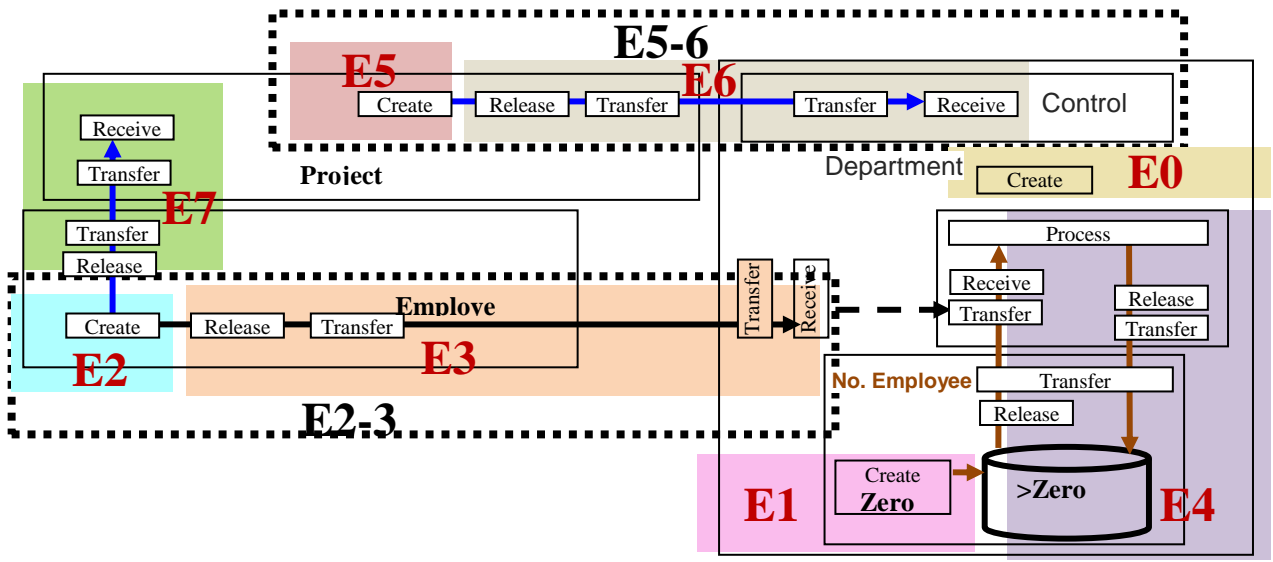

Fig. 22 High-level events E2-3 and E5-6.

Rutle et al. [18] did not mention removing an employee from a department; however, the removed employee had previously caused incrementing of the number of employees, thus, this number is always greater than zero.

**Constraint 4**. A department may control no or many projects.

This rule can be specified by the high-level event E5-6, as shown in the dotted box in Fig. 22, in a manner similar to that described in the previous case of E2-3. The controlling department is specified when creating a project; hence, a department may not have a project to control.

**Constraint 5**. A project must be controlled by at least one department.

The event E2-3 guarantees this rule.

**Constraint 6**. An employee enrolled in a project must work in the controlling department.

Fig. 23 shows the event E2-3-5-6-7, which is indicated by a dotted boundary. It indicates, for employee x, department y, and Project z, that x is an employee (E2) in department y (E3) enrolled in project z (E7), and project z (E5) is controlled by department y (E6). In other words, if x is an employee in department y and enrolled in project z, then z is controlled by department y.

**Constraint 7**. A set of employees working for a controlling department must not be enrolled in the same controlled project more than once.

This rule is applied to the same high-level event, E2-3-5-6-7, as in the previous rule; however, it is expressed in the behavior model as shown in Fig. 24. In Fig. 24, we need a second level language that expresses E5 (z) where z is a specific project. Then, ¬(E2-3-5(z)-6-7) denotes the end of E2-3-5(z)-6-7, i.e., the employee is no longer enrolled in project z. Assuming the employee is no longer enrolled in project z, then the event E2-3-5(z)-6-7 cannot (cross on the chronology arrow) occur again.

## 6. Conclusions

In this paper, we have expressed constraints diagrammatically using the thinging machine (TM) through examples taken from the UML/OCL literature. This contributes to greater understanding of the notion of constraint in conceptual modeling. It also demonstrates the expressiveness and limitation of the TM. We can conclude that the TM provides as viable a diagrammatic tool as constraints language. Further research will explore this matter further by experimenting with different types of constraints.

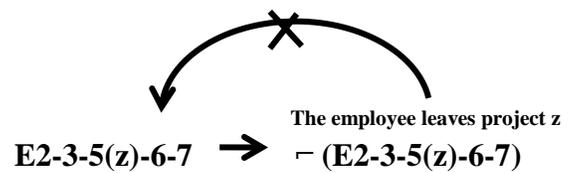

Fig. 24 The behavior that expresses Constraint 7.

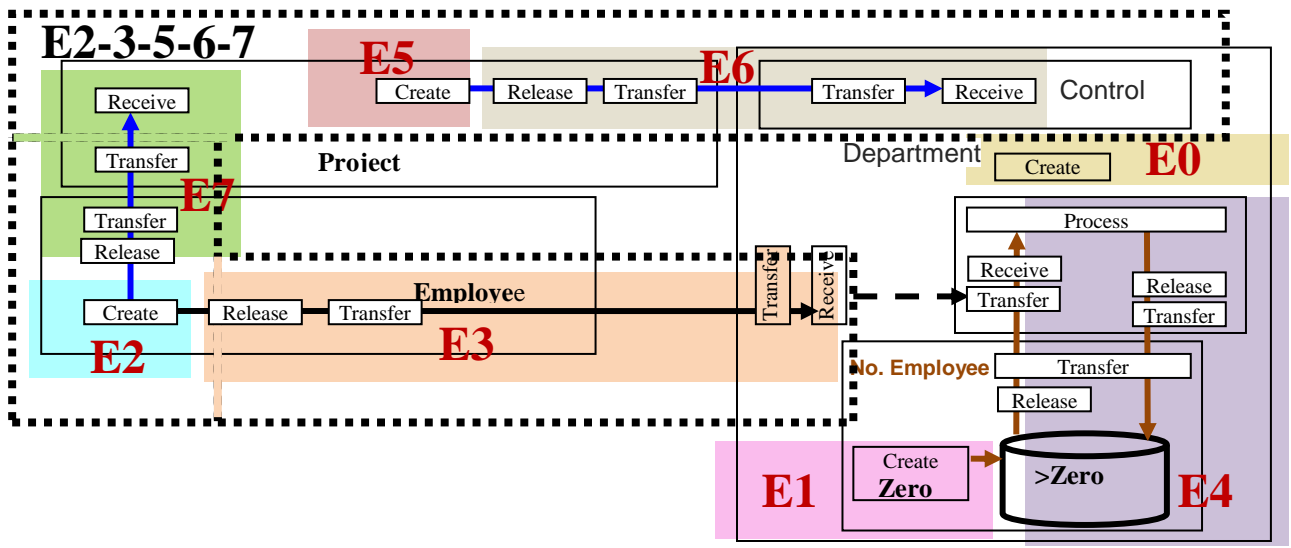

Fig. 23 An Event E2-3-5-6-7.

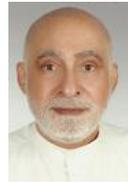

**Sabah S. Al-Fedaghi** is an associate professor in the Department of Computer Engineering at Kuwait University. He holds an MS and a PhD from the Department of Electrical Engineering and Computer Science, Northwestern University, Evanston, Illinois, and a BS from Arizona State University. He has published many journal articles and papers in conferences on software engineering, database systems, information ethics, privacy, and security. He headed the Electrical and Computer Engineering Department (1991–1994) and the Computer Engineering Department (2000–2007). He previously worked as a programmer at the Kuwait Oil Company. Dr. Al-Fedaghi has retired from the services of Kuwait University on June 2021. He is currently (Fall 2021/2022) seconded to teach in the department of computer engineering, Kuwait University.